\documentclass[a4,aps,amsmath,floatfix,nofootinbib,10pt]{revtex4} 
\usepackage{graphicx} \usepackage{color} \usepackage{enumerate} 
\newcommand{\be}{\begin{equation}} \newcommand{\ee}{\end{equation}} 
\newcommand{\ba}{\begin{array}} \newcommand{\ea}{\end{array}} 
\newcommand{\bea}{\begin{eqnarray}} \newcommand{\eea}{\end{eqnarray}} 
\newcommand{\bdm}{\begin{displaymath}} 
\newcommand{\edm}{\end{displaymath}}

\newcommand{\erf}{\operatorname{erf}} \begin{document}


\title{Criteria for Infinite Avalanches in Zero-Temperature 
Nonequilibrium Random-Field Ising Model on a Bethe Lattice}

\author{Prabodh Shukla and Diana Thongjaomayum}

\affiliation{Physics Department \\ North Eastern Hill University \\ 
Shillong-793 022, India}

\begin{abstract}

We present general criteria for the occurrence of infinite avalanches 
and critical hysteresis in the zero-temperature nonequilibrium 
random-field Ising model on a Bethe lattice. Drawing upon extant 
results as well as a new result on a dilute four-coordinated ($z=4$) 
lattice, we show that diverging avalanches can occur if an arbitrarily 
small fraction of sites on a spanning cluster have connectivity $z \ge 
4$.

\end{abstract}

\maketitle

\section{Introduction}

Zero-temperature nonequilibrium random-field Ising model has been used 
extensively to understand hysteresis in systems with quenched 
disorder~\cite{sethna1, sethna2,sethna3,vives,illa,rosinberg,perez, 
spasojevic}. Analysis and simulations of the model reveal that 
character of hysteresis loop depends upon several factors including 
probability distribution of the random-field, dimensionality and 
connectivity of the lattice. A common choice for the random-field 
distribution is a Gaussian with average zero and standard deviation 
$\sigma$. In some cases depending upon the dimensionality and 
connectivity of the lattice, there exists a critical value 
$\sigma=\sigma_c$ such that each half of the hysteresis loop has a 
critical point characterized by diverging susceptibility of the system. 
For $\sigma < \sigma_c$, there is a discontinuity in each half of the 
loop arising from a massive flipping of spins triggered by an 
infinitesimal change in the applied field, i.e. an infinite avalanche. 
Infinite avalanches and a critical point go hand in hand in the present 
model. Infinite avalanches end in a critical point as $\sigma$ is 
increased to its critical value. While the critical behavior in the 
vicinity of $\sigma_c$ has been investigated extensively, the criteria 
for the existence of $\sigma_c$, its dependence on the dimensionality 
and connectivity of the lattice has received less attention. As 
discussed below, general conditions for the occurrence of infinite 
avalanches in this simple model remain unclear so far. This brief 
report addresses this issue drawing upon extant results as well as a 
new result presented here.

\section{The model, extant results, and related issues}

We begin by describing the model briefly, listing known results and 
related issues. The Hamiltonian of the model is, \bdm H=-J\sum_{i,j}s_i 
s_j-h\sum_i s_i -\sum_i h_i s_i \edm

Here $s_i=\pm1$ is an Ising spin at site $i$, $h$ is a uniform applied 
field, and $h_i$ is a quenched field drawn from a Gaussian distribution 
of mean zero and standard deviation $\sigma$; $J$ is ferromagnetic 
coupling between nearest neighbors on a lattice. The applied field $h$ 
is assumed to vary infinitely slowly, and the dynamics of the model is 
taken to be the adiabatic zero-temperature single-spin-flip Glauber 
dynamics. At each value of $h$, spins in the system are flipped as 
needed till each spin $s_i$ is aligned along the net local field 
$\ell_i$ at its site; $\ell_i = J \sum_{j \ne i} s_j + h +h_i$.  We 
start at $h=-\infty$ with the stable state of the system having all 
spins down, and the magnetization per site $m(h)=-1$. Now $h$ is 
increased till some spin flips up, say at $h=\tilde{h}$. A spin that 
flips up increases the field on its neighbors by $2J$, and some of them 
may flip up and so on. Thus a spin flipping up may initiate an 
avalanche of flipped up spins. Sites in an avalanche lie on a connected 
cluster whose size equals the size of the avalanche. Each avalanche of 
size $s$ increases $m(h)$ by an amount $2s/N$ where N is the size of 
the system. When an avalanche is finished, the field $h$ is increased 
again till the next avalanche occurs. This process is continued till 
all spins are up. As $h$ is ramped up from $h=-\infty$ to $h=\infty$, 
$m(h)$ increases in tiny irregular steps separated by random quiescent 
intervals along the applied field. This is Barkhausen noise but not the 
main concern of the present paper. Our concern is with a discontinuity 
in $m(h)$ in the thermodynamic limit, i.e. a macroscopic avalanche of 
the order of $N$ and the criteria for its occurrence.


What we know so far is that if a macroscopic avalanche occurs, it 
occurs for $\sigma < \sigma_c$ where $\sigma_c$ is a critical value. If 
there is no $\sigma_c$, there is no avalanche. Also, there is only one 
avalanche, say at $h=\tilde{h}$. As $\sigma \rightarrow \sigma_c$, 
$\tilde{h}$ decreases and so does the size of the avalanche. The size 
goes to zero at $\sigma = \sigma_c$, $h=\tilde{h}_c$ but fluctuations 
at this point are anomalously large. This is a nonequilibrium critical 
point with behavior similar to that of an equilibrium Ising model at 
its critical temperature $T_c$. Indeed, $\sigma_c$ plays a role 
analogous to $T_c$. It is not clear why this should be so because $T_c$ 
takes into account thermal relaxation of all states of the system, but 
$\sigma_c$ is based only on one initial state (all spins down) and its 
zero-temperature Glauber dynamics. The question we ask is whether $T_c$ 
and $\sigma_c$ are determined by similar criteria. Our minimal model is 
characterized by a small set of parameters: $J, h, \sigma$, and two 
implicit parameters $d$ and $z$ denoting the dimension and coordination 
number of the lattice. $J$ sets the energy scale, $h$ and $\sigma$ are 
used as tuning parameters to locate a critical point if it exists. This 
means that the existence of a critical point $(\sigma_c, \tilde{h}_c)$ 
must depend on $d$, or $z$, or both. For thermal model, the existence 
of $T_c$ is decided by $d$ alone which should be above the lower 
critical dimension $d_\ell$ for the system; $d_\ell=1$ for the pure 
Ising model, $d_\ell=2$ for the random-field Ising model 
~\cite{imry-ma}. For $d > d_\ell$, the temperature-driven critical 
behavior does not depend on $z$. For example, if there is a $T_c$ on a 
square lattice ($d=2,z=4$), there is also a $T_c$ on the honeycomb 
lattice ($d=2,z=3$). One may expect the same for the disorder-driven 
critical behavior at $T=0$ on grounds that both types of critical 
phenomena are caused by a diverging length; diverging correlation 
length in one case and a diverging avalanche in the other. However, 
there is a $\sigma_c$ on the square lattice~\cite{spasojevic} and the 
triangular lattice ~\cite{diana} but no $\sigma_c$ on the honeycomb 
lattice~\cite{sabhapandit}. We would like to understand why?

The coordination number $z=4$ has a special connection with $\sigma_c$. 
The square lattice ($z=4$) is commonly used for studying the behavior 
of a model in $d=2$. Numerical efforts to find $\sigma_c$ for the 
square lattice were inconclusive initially. This was thought to have a 
bearing on $d_\ell$ for the random-field Ising model which was in 
question initially. Eventually theory settled $d_\ell=2$, and numerical 
work on large systems indicated the presence of $\sigma_c$ on the 
square lattice. However, as mentioned above there appear to be other 
factors beside $d_\ell$ that determine $\sigma_c$. An analytic solution 
of the model on a Bethe lattice of integer coordination number $z$ 
shows that infinite avalanches and critical phenomena occur only if $z 
\ge 4$ ~\cite{dhar}. This is surprising because in all other cases, as 
far as we know, the critical behavior of an Ising model on a Bethe 
lattice does not depend upon $z$ if $z>2$. Evidently, there is not a 
very simple and clear physical reason for the absence of infinite 
avalanches on a $z=3$ Bethe lattice.  It has been explained by mapping 
the problem to a branching process in population dynamics 
~\cite{handford}. The unusual dependence of $\sigma_c$ on the 
coordination number $z$ of a Bethe lattice is seen on periodic lattices 
as well. For example, an infinite avalanche does not occur on any 
periodic lattice with $z=3$ irrespective of the dimension $d$ of the 
space in which the lattice is embedded ~\cite{sabhapandit}.

One way to gain insight into the effect of $z$ on $\sigma_c$ is to 
study the model on lattices whose average coordination number $z_{av}$ 
varies continuously between integer values. With this in mind infinite 
avalanches were studied on a dilute triangular lattice ~\cite{kurbah}. 
The study suggested that infinite avalanches occur when $z_{av} \ge 4$ 
but did not rule out a lower value. Next, the problem was studied on a 
Bethe lattice of a mixed coordination number such that a fraction $c_4$ 
of the sites had $z=4$ and the remaining fraction $1-c_4$ had $z=3$. An 
exact solution was obtained and verified by numerical simulations. The 
result turned out to be somewhat surprising. Infinite avalanches can 
occur in the entire range $0<c_4\le1$ if $\sigma < \sigma_c$ where 
$\sigma_c \rightarrow 0$ continuously as $c_4 \rightarrow 0$ 
~\cite{shukla}. This suggests that the presence of a small fraction of 
$z=4$ sites suffices to produce an infinite avalanche. However, the 
physical reason for this is not clear. 

\section{Criticality on a dilute $z=4$ Bethe lattice}

A dilute $z=4$ Bethe lattice with only a fraction $c$ of sites occupied 
by spins provides another example of interest that can be solved 
analytically. On such a lattice, there are sites with coordination 
numbers $z=0,1,2,3,4$. This problem was studied earlier in the limit $c 
\rightarrow 1$, and $c \rightarrow 0$ to show how a tiny fraction of 
magnetic grains in geological rocks transforms familiar hysteresis loops 
into wasp-waisted loops~\cite{kharwanlang}. In the following, we revisit 
this problem in the regime of moderate $c$ to examine the dependence of 
$\sigma_c$ on $c$. A key quantity is the conditional probability 
$Q^*(h,\sigma)$ that a nearest neighbor of an occupied site in the deep 
interior of a Cayley tree (the central site) is down at $h$ before the 
central site is relaxed. The dynamics of the model is abelian i.e. the 
same final state is reached irrespective of the order in which the sites 
are relaxed. We start with all spins down on a Cayley tree and relax 
them in the following order: first we relax spins on the surface, then 
move towards the center relaxing spins on one level at a time. This 
amounts to calculating $Q^n(h,\sigma)$ for increasing $n$ where 
$Q^n(h,\sigma)$ is the probability that a site on level $n$ is down 
before its neighbor at level $n+1$ is relaxed. Let us take the surface 
to be at level $0$. A spin on the surface experiences a quenched random 
field $h_i$, an external field $h$, and a field $-J$ from the unrelaxed 
neighbor at level 1. When relaxed, it may flip up or stay down depending 
on the value of the net field $h_i+h-J$ on it. The probability 
$Q^0(h,\sigma)$ that it stays down is the probability that $h_i+h-J \le 
0$. Taking into account the Gaussian distribution of $h_i$, we get 
$Q^0(h,\sigma)=0.5 [1 + \erf{ \{(J-h)/\sqrt{2 \sigma^2}\}}]$. An 
equation for $ Q^n(h,\sigma)$, $n > 1 $, is obtained similarly if we 
keep in mind that each spin to be relaxed at level $n$ has one unrelaxed 
neighbor at level $n+1$ and $z-1$ relaxed neighbors at level $n-1$. We 
get,

\be Q^n(h,\sigma)=\sum_{m=0}^{z-1}[Q^{n-1}(h,\sigma)]^m [1 - 
Q^{n-1}(h,\sigma)]^{z-1-m} q_{z,m+1}(h,\sigma) \ee

Here $q_{z,m}(h,\sigma)$ is the probability that a $z$-coordinated spin with 
$m$ neighbors down is down at applied field $h$. 

\bdm q_{z,m}(h,\sigma)=\frac{1}{\sqrt{2\pi\sigma^2}} \int_{-\infty}^{(2 
m-z)J-h} e^{\frac{-h_i^2}{2\sigma^2}} dh_i = \frac{1}{2} \left[1 + 
\erf{\left \{\frac{(2m-z)J-h}{\sqrt{2 \sigma^2}}\right\}}\right]; 
\hspace{.25cm} (m=0,\ldots,z) \edm

The fixed-point $Q^*(h,\sigma)$ is given by, $Q^*(h,\sigma)= \lim_{ 
n\rightarrow \infty} Q^n(h,\sigma)$. We find $Q^*(h,\sigma) > 1/2$ if 
$h < J$, $Q^*(h,\sigma) < 1/2$ if $h > J$. There is a discontinuity in 
$Q^*(h,\sigma)$ at $h=J$ and therefore an infinite avalanche in the 
system if $\sigma < \sigma_c$. The discontinuity deceases in size with 
increasing $\sigma$ and vanishes at $\sigma=\sigma_c$. At $\sigma_c$, 
the two solutions at $h=J$ merge into $Q^*(J,\sigma) =1/2$. The 
equation determining $\sigma_c$ is,

\bdm A (A+4B)=0. \edm where \bdm A=c^3 \{1 + q_{4,4}(J,\sigma)- 3 
q_{2,2}(J,\sigma) \} \mbox{ and } B=1- \{c^3 q_{4,4}(J,\sigma)+3 c^2 
(1-c) q_{3,3}(J,\sigma)+3 c (1-c)^2 q_{2,2}(J,\sigma)+(1-c)^3 
q_{1,1}(J,\sigma)\}\edm

The factor $A$ is negative for all values of $\sigma$ of interest. 
Therefore $\sigma_c$ is effectively determined by the equation $A+4B=0$. 
No real positive value of $\sigma_c$ satisfies this equation if $c$ is 
less than a critical value $c_{min}$. Numerically, $c_{min}\approx 
0.557$. The exact value (argument to be presented below) is $c_{min} = 
2^{1/3} / (1+2^{1/3}) \approx 0.5575$. For $c > c_{min}$, $\sigma_c$ 
increases continuously with increasing $c$ starting from $\sigma_c=0$ at 
$c=c_{min}$. The increase is remarkably steep in a narrow region 
adjacent to $c_{min}$. We may designate this region as the critical 
region. The width of the critical region is very small but the increase 
of $\sigma_c$ in this region is substantial. Thus a plot of $\sigma_c$ 
vs. $c$ appears almost vertical at $c=c_{min}$. Theoretically, the slope 
of $\sigma_c$ vs. $c$ curve is infinite at $c_{min}$. It gradually 
decreases as one moves away from $c_{min}$ but remains very large over 
the entire critical region. Figure (1) shows a plot of $\sigma_c$ vs. 
$c$; $\sigma_c$ appears to rise vertically from $\sigma_c=0$ to 
$\sigma_c\approx0.275$ i.e. all values in the range $0 < \sigma_c < 
0.275$ satisfy the equation at $c \approx 0.557$. Thereafter $\sigma_c$ 
increases more gradually. At $c=1$ we recover the known result 
$\sigma_c=1.781$ for the undiluted $z=4$ Bethe lattice. The data plotted 
in figure (1) was obtained using a standard numerical recipe for 
evaluating error functions in the expression for $\sigma_c$. The error 
in using this recipe is $\epsilon \le 10^{-7}$. Within this error, 
$\sigma_c$ at $c_{min}$ rises vertically as shown in figure (1). 
However, if one uses another tool ( Mathematica ) to calculate the error 
functions with a greater precision ($\epsilon \le 10^{-16}$), the 
vertical portion of the $\sigma_c$ vs. $c$ curve is replaced by a curve 
that bends slightly to the right in the narrow critical region; 
$\sigma_c$ increases continuously from 0.0 to 0.307275 as $c$ increases 
from $c_{min}\approx 0.5575$ to $c=0.558$ ~\cite{referee}. This 
continuous but sharp increase in a narrow region would also appear to be 
nearly vertical when plotted on the scale of figure (1). The important 
point is that mathematical tools with higher precision as well as 
theoretical analysis agree that $\sigma_c$ is a continuous, monotonic, 
but very steeply increasing function of $c$ immediately above the 
threshold $c=c_{min}$. The situation brings to mind some (not so well 
understood) transitions in liquid crystals where the order parameter 
appears to jump discontinuously as in a first-order transition but the 
entropy changes continuously as in a second-order transition. In order 
to confirm this sharp change at $c \approx 0.557$, we performed 
simulations for $m(h,\sigma)$ at $\sigma=0.4$ for $c=0.55$ and $c=0.57$. 
Theory predicts a discontinuity in $m(h,\sigma=0.4)$ for $c=0.57$ but no 
discontinuity for $c=0.55$. This is what we observed in the simulations. 
Figure (2) shows the closeness between the numerical and corresponding 
theoretical results. The numerical results are obtained on a random 
graph rather than a Cayley tree in order to eliminate large surface 
effects. The initial state of the random graph is taken as all spins 
down. These simulations match the theoretical result on a Cayley tree if 
the surface spins are kept down, rather than relaxed at $h$. This 
procedure does not alter $\sigma_c$, but shifts the discontinuity from 
$h=J$ to to $h > J$. The discontinuity moves closer to $h=J$ as $\sigma$ 
increases, and vanishes at $h=J$ as $\sigma \rightarrow \sigma_c$. If 
the surface of the Cayley tree is relaxed at $h$ rather than held in a 
fixed state, then the discontinuity occurs at $h=J$ only.

\section{General criteria for infinite avalanches}

To recapitulate, (i) infinite avalanches occur if $z \ge 4$ and $\sigma 
< \sigma_c(z)$ but do not occur if $z=2, 3$ (ii) on a lattice with 
$c:1-c$ mixture of $z=4$ and $z=3$ sites, infinite avalanches occur for 
all $c$ ($0 < c \le 1$) if $\sigma < \sigma_c(c)$, (iii) on a $z=4$ 
lattice with a fraction $c$ of sites occupied, infinite avalanches occur 
if $c > 0.557$ and $\sigma < \sigma_c(c)$. The reason why infinite 
avalanches do not occur for large $\sigma$ irrespective of other 
considerations is simple. Spins tend to flip up independently in the 
presence of large disorder, hence no infinite avalanche. Results (ii) 
and (iii) are puzzling at first sight; (ii) suggests that an arbitrarily 
small fraction of $z=4$ sites is sufficient to cause an infinite 
avalanche but (iii) contradicts it because nearly $5\%$ sites have $z=4$ 
at $c \approx 0.557$. This requires further discussion. First, we look 
at the reason for (i). Absence of an infinite avalanche means 
$Q^*(h,\sigma)$ is continuous at $h=J$. It is easy to verify that 
$Q^*(J,\sigma) = 1/2$ is a fixed point irrespective of $z$ and $\sigma$. 
The absence or presence of an infinite avalanche depends on the 
stability of this fixed point. If $Q^*(J,\sigma) = 1/2$ is stable, there 
is no discontinuity in $Q^*(h,\sigma)$ at $h=J$. An unstable 
$Q^*(J,\sigma) = 1/2$ splits into two stable fixed points, one larger 
and the other smaller than $1/2$ at $h=J$. Consequently the system jumps 
from a small magnetization state to a large magnetization via an 
infinite avalanche. The stability of $Q^*(J,\sigma) = 1/2$ is examined 
by turning down a small fraction of up sites on the surface of the 
Cayley tree and examining its effect on the next layer of sites. In 
other words, we increase the fraction of down sites on the surface from 
$1/2$ to $1/2 + \delta Q^0$, and calculate the fraction $1/2 + \delta 
Q^1$ of down sites on the layer next to the surface. Focus on a set of 
$z-1$ sites on the surface which have a common neighbor, say $B$ at the 
higher level. Consider the case when at least one of the $z-1$ sites, 
say $A$ is up and $B$ is also up. Now if $A$ is turned down, the local 
field at $B$ gets reduced by $2J$. The probability that $B$ will turn 
down as a result of it is given by $\delta Q^1(J,\sigma) = B_z \delta 
Q^0(J,\sigma)$. Using $q_{z,k}=q_{z,k}(h,\sigma)$ defined earlier, we 
obtain

\bdm B_z = (z-1)\frac{1}{2^{z-2}} \sum_{m=0}^{z-2}{{z-2}\choose{m}} 
(q_{z,m+2}-q_{z,m+1}) \edm

Above equation is understood as follows: site $A$ can be chosen in 
$z-1$ ways, remaining $z-2$ sites are down with probability 
$\frac{1}{2}$, $(q_{z,m+2}-q_{z,m+1})$ is the probability that site $B$ 
is up if $m+1$ of its neighbors are down but flips down if $m+2$ 
neighbors are down, $q_{z,k}$ is the probability that a $z$-coordinated 
spin is down if $k$ of its neighbors are down. The quantities $B_2 = 
q_{2,2}-q_{2,1}$, $B_3 = q_{3,3}-q_{3,1}$, and $B_4 = q_{4,4} + q_{4,3} 
- q_{4,2} - q_{4,1}$ are of special interest. $B_2$ and $B_3$ are less 
than unity for $\sigma > 0$; as $\sigma \rightarrow 0$, $B_2 
\rightarrow 1$ and also $B_3 \rightarrow 1$. Hence the fixed point 
$Q^*(J,\sigma) = 1/2$ is stable and the possibility of an infinite 
avalanche is ruled out on a $z=2$ or a $z=3$ lattice. At $h=J$, $B_4$ 
simplifies to $B_4=\frac{3}{2}(q_{4,4}-q_{4,2})$; $B_4 \rightarrow 3/2$ 
as $\sigma \rightarrow 0$. It decreases continuously with increasing 
$\sigma$; $B_4 \rightarrow 1$ as $\sigma \rightarrow \sigma_c \approx 
1.781$. Thus $Q^*(J,\sigma) = 1/2$ is unstable on a $z=4$ lattice if 
$\sigma < \sigma_c$ and consequently there is an infinite avalanche in 
this case. This also confirms that $\sigma_c$ obtained from considering 
the stability of $Q^*(J,\sigma) = 1/2$ is the same as obtained from 
requiring two roots of the fixed point equation to merge into each 
other. Figure (3) shows the initial value $Q^0(h,\sigma = J)$ and 
corresponding fixed-point value $Q^*(h,\sigma = J)$ in the neighborhood 
of $h=J$ for $z=2, 3,$ and $4$. For $h < J$, the bottom line represents 
surface $Q^0(h,J)$; higher curves show fixed points $Q^*(h,J)$ for 
$z=2, 3,$ and $4$ respectively. The relative position of $Q^0(h,J)$ and 
$Q^*(h,J)$ gets reversed for $h > J$; $Q^*(h,J) > Q^0(h,J)$ if $h < J$, 
but $Q^*(h,J) < Q^0(h,J)$ if $h > J$. $Q^0(h,J)$ is of course 
continuous at $h=J$ and $Q^0(J,J)=0.5$; $Q^*(h,J)$ is continuous at 
$h=J$ if $z \le 3$ but discontinuous if $z=4$. Figure (4) shows the 
growth for $z=4$, and decay for $z=2, 3$ of a small perturbation 
$\delta Q^0(h = J,\sigma = J)$ under successive iterations.

Next, we turn our attention to the dilute $z=4$ lattice. In this case 
as well, $Q^*(J,\sigma) = 1/2$ is a fixed point. This fixed point must 
be unstable in the limit $\sigma \rightarrow 0$ if there is to be an 
infinite avalanche. In the limit $\sigma \rightarrow 0$, $B_2 
\rightarrow 1$, $B_3 \rightarrow 1$, and $B_4 \rightarrow 3/2$. 
Consider a perturbation $\delta Q^0(J,0)$ to the fixed point 
$Q^*(J,\sigma) = 1/2$. As we move from the surface of the tree towards 
its center, a $z=4$ site increases the perturbation by a factor $3/2$, 
but $z=3$ and $z=2$ sites keep it unchanged. The $z=1$ sites produce a 
new effect on the dilute lattice. They break the continuity of the path 
from the surface to the center. In our algorithm for relaxing sites, it 
is assumed that one of the neighbors of the site being relaxed, the one 
at a higher level, is present and unrelaxed. The bond with this 
neighbor ensures connection between adjacent levels of the tree. A 
$z=1$ site breaks this connection with probability $\frac{3}{4} z_1$, 
where $z_1$ is the fraction of sites with one nearest neighbor only. If 
$z_4$ is the fraction of sites with $4$ neighbors, then at every level 
of relaxation of the lattice, the perturbation is boosted with the 
probability $\frac{3}{2} z_4$, and terminated with probability 
$\frac{3}{4} z_1$. The critical point occurs when the two opposing 
effects balance each other, i.e. $z_1 = 2 z_4$. Using $z_4=c^5$, $z_1=4 
c^2 (1-c)^3$, the critical value of $c$ is given by the equation 
$c^3=2(1-c)^3$, or $c=2^{1/3}/(1+2^{1/3}) \approx 0.5575$. The observed 
infinite avalanche on a mixed lattice with a fraction $c_4$ of $z=4$ 
sites and $1-c_4$ of $z=3$ sites for $c_4 > 0$ is also understood in 
this light. The path from the surface to the center is never broken on 
the mixed lattice, and therefore an arbitrarily small presence of $z_4$ 
sites creates a gap in $Q^*(J,\sigma)$ in the deep interior of the 
tree.

\section{Conclusion}

To conclude, we have presented general criteria for the occurrence of 
infinite avalanches in the zero-temperature nonequilibrium random-field 
Ising model on a Bethe lattice. We find that infinite avalanches occur 
when all of the following conditions are fulfilled: (i) $\sigma$ is 
sufficiently small, (ii) there is a spanning cluster of occupied sites 
on the lattice, and (iii) the spanning cluster has a fraction of sites, 
even an arbitrarily small fraction, with connectivity $z \ge 4$. We 
have explained the reason for these conditions. The presence of an 
infinite avalanche on a mixed coordination lattice ($z=3$ or $4$) with 
an arbitrarily small fraction of $z=4$ sites, and its absence on a 
dilute $z=4$ lattice in a certain regime of dilution is now easily 
understood.

Our analysis also shows that disorder in the form of dilution of 
magnetic ions on a lattice affects hysteresis differently from disorder 
in the form of on-site random-fields. This is important because 
positional disorder in the form of vacancies is quite common in 
materials. We find a peculiar geometry driven transition near 
$c=0.5575$ on a dilute $z=4$ Bethe lattice. Similar behavior may be 
expected for $z>4$ as well. Infinite avalanches vanish at this critical 
point continuously, but the slope of the continuous curve is nearly 
infinite. It appears as a first-order jump in the order parameter for 
all practical purposes. Bethe lattices often approximate real systems 
reasonably well. So this feature of the model may be observable in 
appropriate hysteresis experiments and useful in understanding other 
weakly first-order phase transitions as well.

Finally, we wish to end with a caution. We have made a case that a 
lower critical coordination number rather than a lower critical 
dimension determines critical hysteresis. Our suggestion is based on 
exact results on Bethe lattices and simulations on some periodic 
lattices. It conflicts with a widely accepted view in statistical 
physics community in favor of a lower critical dimension. Further work 
may be required to settle this issue but we mention two factors that 
may invalidate our suggestion. Results on Bethe lattices are 
essentially mean field results and do not necessarily have a bearing on 
criticality on periodic lattices in finite dimensions. Secondly, subtle 
corrections to scaling may explain the extant numerical results on 
periodic lattices without doing away with the importance of a lower 
critical dimension.

\begin{figure}[p] 
\includegraphics[width=0.75\textwidth,angle=0]{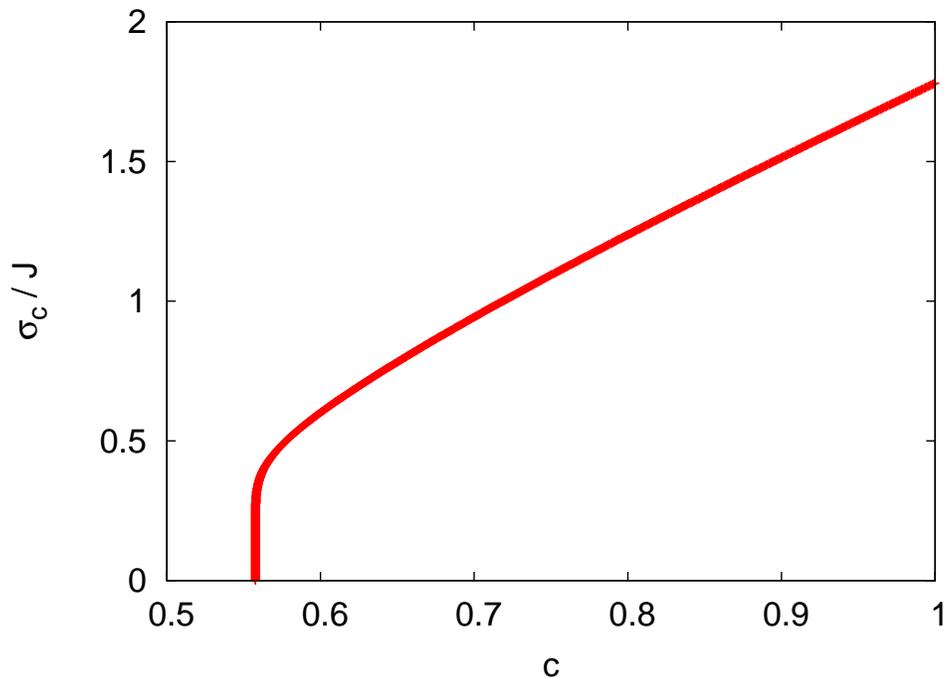} 
\caption { Critical value of the standard deviation of the quenched random-field 
$\sigma_c$ on a $4$-coordinated random graph with a fractional occupancy $c$ of its 
sites.  The magnetization $m(h,\sigma)$ has a discontinuity if $\sigma < \sigma_c$. 
There is an almost vertical drop in $\sigma_c$ at $c=0.557$ approximately. } 
\label{fig1} \end{figure}

\begin{figure}[p] 
\includegraphics[width=0.75\textwidth,angle=0]{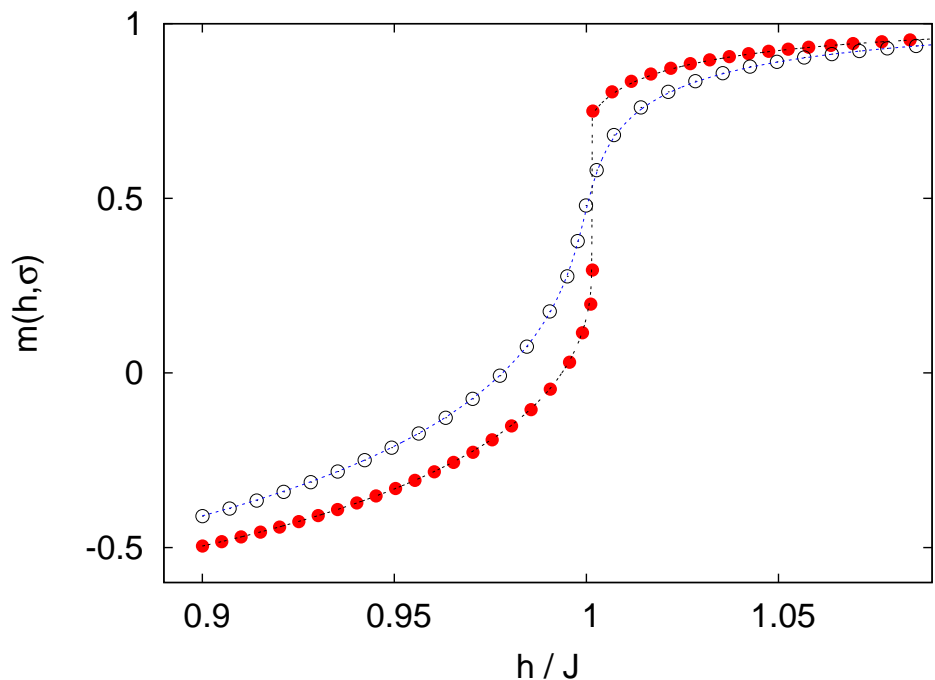} \caption 
{Magnetization $m(h,\sigma)$ in increasing field $h$ on a 
$4$-coordinated dilute random graph for $c=0.55$ (open circles) and 
$c=0.57$ (filled circles) where $c$ is the fraction of occupied sites 
on the graph. The quenched random field on occupied sites has mean 
value equal to zero, and standard deviation $\sigma=0.40$. Theoretical 
predictions are superimposed on the respective simulations (a single 
run on $N=10^7$ graph) and fit them quite well. The magnetization is 
smooth for $c=0.55$ and has a jump for $c=0.57$ as predicted by the 
theory.} 
\label{fig2} \end{figure}

\begin{figure}[p] \includegraphics[width=0.75\textwidth,angle=0]{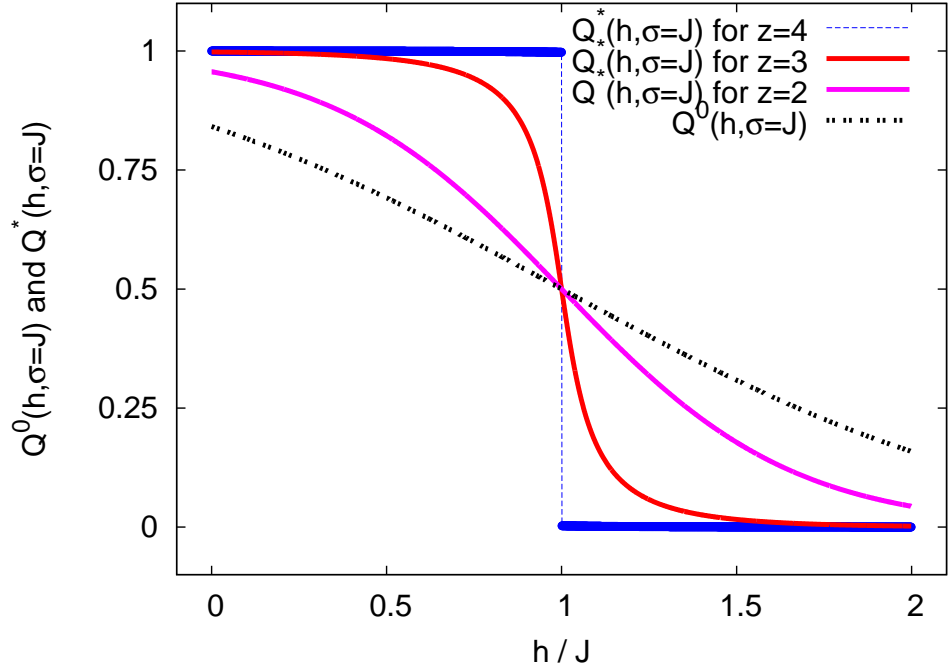} 
\caption{The figure shows $Q^0(h,\sigma=J)$ and $Q^*(h,\sigma=J)$ in increasing 
field $h$ (lower half of the hysteresis loop) on a Cayley tree of coordination 
number $z=2, 3, 4$. The broken black line shows $Q^0(h,\sigma=J)$ which is common 
to all $z$ because a surface site has only one neighbor irrespective of $z$; 
$Q^0(h,\sigma=J)$ decreases continuously with increasing $h$ and passes through the 
point $Q^0(h,\sigma=J)=0.5$ at $h=J$. $Q^*(h,\sigma=J)$ for $z=2$ (pink curve 
closest to the black broken line) and $z=3$ (red curve next closest to the black 
broken line) behave similarly; both decrease continuously and pass through 
$Q^*(h,\sigma=J)=0.5$ at $h=J$. However, as we go from $z=2$ to $z=3$, 
$Q^*(h,\sigma=J)$ becomes steeper at $h=J$, and generally moves farther away from 
$Q^0(h,\sigma=J)$. For $z\ge4$, $Q^*(h,\sigma=J)$ acquires a discontinuity at $h=J$ 
as shown in the figure by the blue curve.} \label{fig3} \end{figure}

\begin{figure}[p] \includegraphics[width=0.75\textwidth,angle=0]{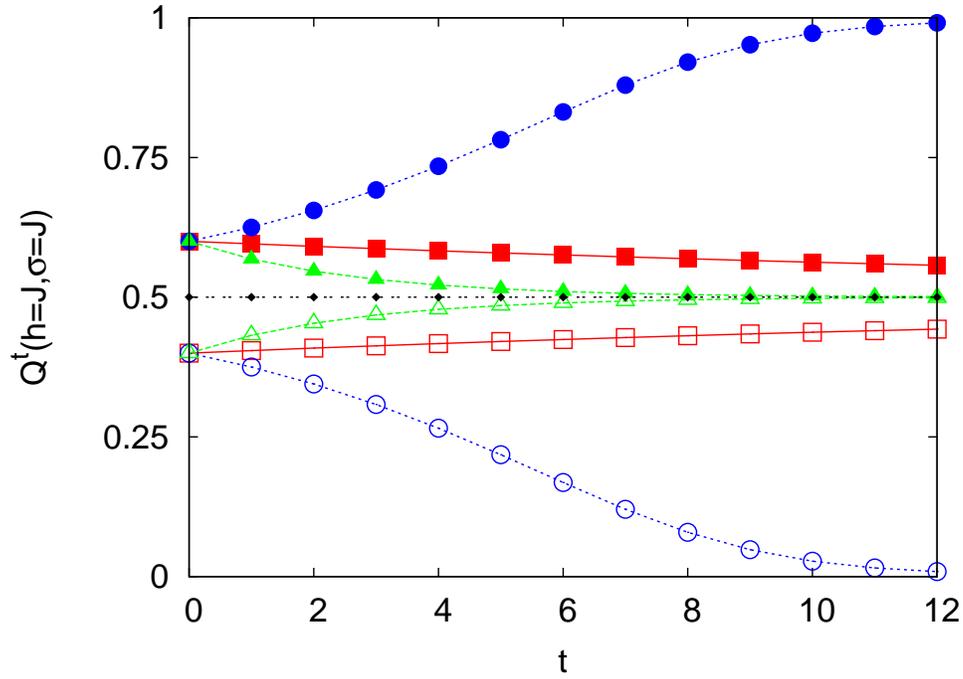} 
\caption{The conditional probability $Q^t(J,J)$ (see text) at $t$ successive levels of 
the Cayley tree starting from the surface ($t=0$). If $Q^0(J,J)=0.5$, $Q^t(J,J)$ 
remains equal to $0.5$ as shown by the horizontal line. A small perturbation $\delta 
Q^0$ added to $Q^0(J,J)$ gradually decreases to zero if $Q^*(J,J)=0.5$ is stable, but 
increases with $t$ if $Q^*(J,J)=0.5$ is unstable; perturbed $Q^t(J,J)$ lies on the 
same side of the horizontal line $Q^*(J,J)=0.5$ as the initial perturbation $\delta 
Q^0$. Figure shows $Q^*(J,J)=0.5$ is stable for $z=2$ (green triangles) and $z=3$ (red 
squares), but unstable for $z=4$ (blue circles). Filled symbols correspond to $\delta 
Q^0 > 0$ while the empty symbols correspond to $\delta Q^0 <0$. The relaxed state on 
the first twelve levels of the tree is shown which suffices to make the trends clear.} 
\label{fig4} \end{figure}

\end{document}